\documentclass{elsart}
\usepackage[]{graphicx,times}

\newcommand{\pbarp}{\overline{p}p}
\newcommand{\KKbar}{K\overline{K}}
\newcommand{\qqbar}{q\overline{q}}
\newcommand{\uubar}{u\overline{u}}
\newcommand{\ddbar}{d\overline{d}}
\newcommand{\ssbar}{s\overline{s}}
\newcommand{\nnbar}{n\overline{n}}
\begin{document}
\begin{frontmatter}

\title{Further  evidence for a large glue component in the
f$_{\bf 0}$(1500) meson}
\author{Claude Amsler\thanksref{label1}}
\address{Physik-Institut der Universit\"at Z\"urich,
Winterthurerstrasse 190, \\CH-8057 Z\"urich, Switzerland}
\thanks[label1]{{\it e-mail address:} Claude.Amsler@cern.ch\\
{\it \hspace*{3mm} url:} http://afuz01.cern.ch/zuerich.html}

\begin{abstract}
We argue that the experimental decay rates of the $f_0(1500)$ meson into two pseudoscalar
mesons and into two photons are incompatible with a quark-antiquark state and that the
$f_0(1710)$ meson is  dominantly an 
$\ssbar$ state. 
\end{abstract}

\begin{keyword}
Glueball, meson, scalar, f0(1500); 
12.38.Qk, 12.39.Mk,14.40.Cs

\end{keyword}
\end{frontmatter}

\section{Introduction}
Several $0^{++}$ resonances were established recently, far too many to be  accommodated
in the ground state scalar nonet \cite{Groom}. Below 1600 MeV, four isospin zero resonances
are required to describe the isoscalar $\pi\pi$ S-wave: the very broad ($\sim$ 600
MeV) $f_0(600)$ (or
$\sigma$), the narrow $f_0(980)$ at the  $\KKbar$ threshold, the  broad ($\sim$ 400 MeV)
$f_0(1370)$, and the comparatively narrow (109 $\pm$ 7 MeV) $f_0(1500)$. The latter, first
observed in $\pbarp$ annihilation at rest, was reported in many other reactions and is among
the best  established unstable mesons. For example, it is observed to decay into $\pi\pi$
\cite{Amsler95f,Bertin97c}, $\KKbar$
\cite{Domb96},
$\eta\eta$ \cite{Amsler95e}, $\eta\eta'$ \cite{Amsler94} and $4\pi$ \cite{Abele01} in $\pbarp$
annihilation at rest,  and in $pp$ central collisions at 450 GeV, into $\pi\pi$ and $\KKbar$
\cite{Barberis99d},
$\eta\eta$ \cite{Barberis00e}, $\eta\eta'$ \cite{Barber} and $4\pi$ \cite{Barberis00c}. It also reported in an
analysis of the
$D_s\to 3\pi$ Dalitz plot \cite{Frabetti} and in the $4\pi$ mass distribution in $J/\psi\to
4\pi$ radiative decays \cite{Bugg}. It was, however, not observed so far in $\gamma\gamma$
collisions \cite{Acciarri,Barate}. 

The small branching ratios for $\KKbar$ decay - compared to $2\pi$ - of  $f_0(1370)$
and $f_0(1500)$ point to both states being compatible with $\uubar + \ddbar$
structures, while only one can be accommodated in a SU(3) nonet. In an earlier work 
\cite{FEC} we suggested that three isoscalar mesons in the 1500 MeV mass region,
a $\uubar+\ddbar$ state, an $\ssbar$ and the ground state scalar glueball predicted by lattice
gauge theories \cite{Michael} mix to produce $f_0(1370)$, $f_0(1500)$ and the at
that time still to be identified third scalar state. In this  scenario, a large ($\geq 50\%$)
gluonic component was suggested in the $f_0(1500)$ wavefunction, while $f_0(1370)$ remained
largely $\uubar+\ddbar$, and the missing state mainly $\ssbar$ . This was inferred (i) from
the apparent absence of
$\KKbar$ signal in earlier bubble chamber exposures suggesting a cancellation between
amplitudes, (ii) from the narrow width of $f_0(1500)$  compared to the other known scalars,
and (iii) from the observation of an isovector $a_0(1450)$ in this mass region. In this model
the ground state ($1^3P_0$) scalar nonet was therefore made of
$a_0(1450)$, $f_0(1370)$, an $\ssbar$ state around 1700 MeV (both with   small glue
components), and  $K_0^*(1430)$, while $f_0(1500)$ was mainly glue. 

On the other hand, the $f_0(980)$ and $a_0(980)$ could be $\KKbar$ molecules or four-quark
states
\cite{Molecules}, or together with $f_0(600)$, meson-meson resonances \cite{Oller}.
Alternative mixing schemes also imbedding quarkonia $f_0(1370)$ and $f_0(1500)$ have
been proposed \cite{Altern,Weingarten}.

The $\KKbar$ decay mode of $f_0(1500)$ was observed meanwhile \cite{Domb96,Barberis99d} and is
somewhat stronger than the upper limit from bubble chamber experiments, but   still much
weaker than the $\pi\pi$ mode. This indicates that $f_0(1500)$, if interpreted as a $\qqbar$
state, can have  only a small $\ssbar$ component. Also, the missing $\ssbar$ scalar
has now been observed: the longstanding controversy on the spin of the $f_J(1710)$ ($J$ = 0
or 2) was lifted in favour of $J=0$ \cite{Barberis99d,Dunwoodie}.  The early amplitude
analysis of former central production data assumed the spin 2
$f_2'(1525)$ around 1500 MeV, but no  $f_0(1500)$. An updated analysis along the lines of ref. \cite{FEC},
but using instead of Crystal Barrel data more recent results from central collisions, leads to 83
\% $\ssbar$ in $f_0(1710)$ and 60 \% $\uubar+\ddbar$ in $f_0(1370)$, while $f_0(1500)$
contains the largest fraction of glue (48 \%) \cite{Kirk}.  

The purpose of this letter is to discuss the impact of these and other new data on the quark
structures of $f_0(1500)$ and $f_0(1710)$. In particular, we will show that for $f_0(1500)$
the $\KKbar$ and $\gamma\gamma$ data do not appear to be consistent with a dominantly
quarkonium state, while the $\KKbar$ data suggest a large $\ssbar$
component in the $f_0(1710)$ wavefunction.

\section{Couplings to two pseudoscalar mesons}
The branching ratios for $f_0(1500)$ decay into $\pi\pi$, $\KKbar$ and $\eta\eta$
have been measured by several groups: Crystal Barrel quotes from a coupled channel analysis
of the $3\pi^0$, $\pi^0\pi^0\eta$ and $\eta\eta\pi^0$ data in $\pbarp$ annihilation at rest
\cite{coupled} and from the
$\pi^0K_LK_L$ channel \cite{Domb96}:
\begin{eqnarray}
\Gamma(\eta\eta)/\Gamma(\pi\pi) & = &   0.157  \pm  0.062, \\ 
\Gamma(\KKbar)/\Gamma(\pi\pi) & = & 0.119  \pm  0.032,
\end{eqnarray}
while WA102 reports in $pp$ central collisions \cite{Barberis00e}
\begin{eqnarray}
\Gamma(\eta\eta)/\Gamma(\pi\pi) & = & 0.18   \pm  0.03,\\
\Gamma(\KKbar)/\Gamma(\pi\pi)  & = & 0.33  \pm  0.07.
\end{eqnarray} 
The $\eta\eta/\pi\pi$ ratio was also measured by Crystal Barrel with 900 MeV/c
antiprotons \cite{Amsler02}: 0.080 $\pm$ 0.033. The 
preliminary results for $\KKbar/\pi\pi$ from the Obelix collaboration is 0.24
$\pm$ 0.02 \cite{Semprini}. For
$\KKbar/\pi\pi$ the agreement between experiments is not particularly
good. However, branching ratios are sensitive to interference effects and are not defined
unambiguously. The apparent discrepancy should therefore not be overemphasized. The
disagreement is perhaps due to the nearby $a_0(1450)$ which absorbs a significant
fraction of the $\KKbar$ rate in Crystal Barrel data, but is not required by the WA102 data,
and reported at the lower mass of 1300 MeV for the Obelix data. In any case, all data indicate
that the coupling of $f_0(1500)$ to $\KKbar$ is small compared to $\pi\pi$.

To quantify this result we recall the formulae to compute the partial width of a
scalar (or tensor) meson decaying into two pseudoscalar mesons $M_1$ and $M_2$ \cite{FEC}
\begin{equation}
\Gamma (M_1M_2) = \gamma^2 (M_1M_2) \times q^{2\ell+1} \times \exp(-q^2/8\beta^2) \ ,
\label{gammas}
\end{equation}
with $\beta$ $\simeq$ 0.5 GeV/c, where $q$ is
the breakup momentum and $\ell$ the relative angular momentum (0 or 2). The couplings
$\gamma^2$ are derived from SU(3)  and read, up to a common multiplicative constant:
\begin{eqnarray}
\gamma^2(\pi\pi) & = & 3 \left [\cos\alpha\right ]^2 \ , \nonumber \\ 
\gamma^2(\KKbar) & = &  \left[ \cos\alpha \left( 1 - \sqrt{2} \tan\alpha\right )\right ]
^2 \ , \nonumber \\
\gamma^2(\eta\eta) & = &  \left[ \cos\alpha \left(\cos^2\phi - \sqrt{2}
\tan\alpha\sin^2\phi\right )\right ] ^2 \ .
\label{f3}
\end{eqnarray}
Here $\alpha$ = 54.7$^\circ + \theta$ and $\phi$ = 54.7$^\circ + \theta_{PS}$. The
angle  $\theta$ is the octet-singlet mixing angle  in the 
nonet of the decaying meson and  $\theta_{PS}$ the mixing angle in the pseudoscalar nonet. We shall use the value 
$\theta_{PS}$ = -- (17.3 $\pm$ 1.8)$^\circ$ 
measured from $\pbarp$ annihilation rates into two pseudoscalar mesons \cite{Pseudo}. 

The wavefunction of the isoscalar $\qqbar$ meson, say $f'$, ist then given by
\begin{equation}
|f'\rangle = \cos\alpha \  |\nnbar\rangle - \sin\alpha \ |\ssbar\rangle \ \ \  {\rm with} \
\ \  |\nnbar\rangle\equiv\frac{\uubar+\ddbar}{\sqrt{2}} \ .
\label{psi}
\end{equation}      

\begin{figure}[htb]
\parbox{140mm}{\mbox{
\includegraphics[width=100mm]{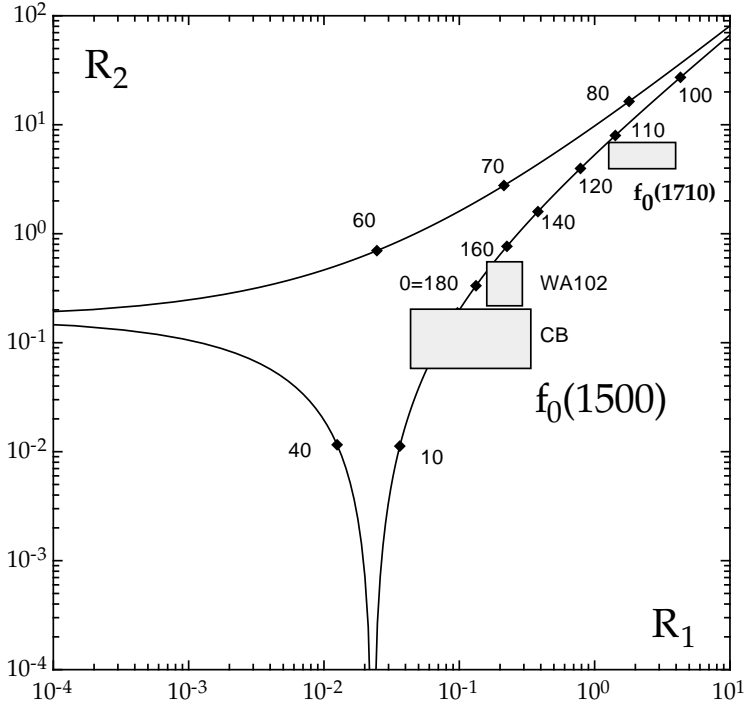}
}\centering}\hfill
\caption[]{Predicted relative coupling $R_2$ = $\gamma^2(\KKbar)/\gamma^2(\pi\pi)$  vs. $R_1$
= $\gamma^2(\eta\eta)/\gamma^2(\pi\pi)$  (curve) compared to data (2$\sigma$  boundaries) from
Crystal Barrel (CB) and  WA102 for the
$f_0(1500)$  and from WA102 for the
$f_0(1710)$. The numbers on the curve indicate  values of the mixing angle
$\alpha$ in degrees.
\label{etakkpipi}}
\end{figure}

Hence for $\alpha$ = 0 the meson is pure $\nnbar$ and for $\alpha$ = 90$^\circ$ 
pure $\ssbar$ (ideal mixing).     These prescriptions provide a very good description e.g.
of the measured tensor meson decays  \cite{FEC}, leading to a tensor mixing
angle $\theta$ 
$\simeq$ 27$^\circ$, in accord with the linear and quadratic mass formulae.  Note that, as
shown in ref.
\cite{FEC}, possible SU(3) breaking effects are small, and are accordingly neglected in our
formulae (\ref{f3}) (hence the parameter $\rho$  in the appendix A of ref. \cite{FEC} is set
to unity).

Let us  calculate the measured ratios $R_1\equiv\gamma^2(\eta\eta)/\gamma^2(\pi\pi)$ and
$R_2\equiv\gamma^2(\KKbar)/\gamma^2(\pi\pi)$ using eqn. (\ref{gammas}). We obtain
\begin{equation} 
R_1 = 0.196 \pm 0.077 \ {\rm and} \  0.227 \pm 0.035 \ ,
\end{equation}
for Crystal Barrel and WA102, respectively, while
\begin{equation} 
R_2 = 0.139 \pm 0.037 \ {\rm and} \ 0.375 \pm 0.082 \ .
\end{equation} 
Note that for a pure $\uubar+\ddbar$ state $R_2$ is expected to be 1/3. Figure \ref{etakkpipi}
shows the predicted dependence of the ratios $R_1$ and
$R_2$ on the angle $\alpha$ for isoscalar mesons and the experimental results for $f_0(1500)$.
We find that
$\alpha
\simeq 0$ (or 180$^\circ$). The $\eta\eta'$ decay rate from Crystal Barrel  is also
consistent with $\alpha \simeq 0$ \cite{RMP}.  If
interpreted as a $\qqbar$ state the meson $f_0(1500)$ is therefore clearly dominantly
$\nnbar$. 

Two comments are in order here. First, this conclusion is not sensitive to a
reasonable departure ($\sim 20 \%$) of the SU(3) breaking $\rho$ from unity. Second, assuming instead a
pseudoscalar mixing angle
$\theta_{PS}$ = --10$^\circ$, as suggested by the quadratic mass formula for the
$0^{-+}$ nonet \cite{Groom}, leads to a very poor agreement of the measured $R_1$ and $R_2$ with SU(3) for
any angle $\alpha$. In this case, it is difficult to reconcile $f_0(1500)$ with a
$\qqbar$ state.

So far, only one experiment, WA102, measured the corresponding ratios of partial widths for the
$f_0(1710)$ meson and reported \cite{Barberis00e}
\begin{eqnarray}
\Gamma(\eta\eta)/\Gamma(\pi\pi) & = &   2.4  \pm  0.6, \\ 
\Gamma(\KKbar)/\Gamma(\pi\pi) & = & 5.0  \pm  0.7,
\end{eqnarray}
leading to the ratios $R_1 = 2.63 \pm 0.66$ and $R_2 = 5.37 \pm 0.75$, which are also shown
in fig. \ref{etakkpipi}. The data are consistent with a state dominantly  $\ssbar$ 
($\sim$ 86\% for $\alpha=112^\circ$). Furthermore, a pure (ideally mixed) $\ssbar$ meson
should not be produced in $\pbarp$ annihilation due to the  OZI rule and, indeed, Crystal
Barrel does not observe $f_0(1710)$ in $\pbarp$ annihilation at 900 MeV/c, while $f_0(1500)$
is produced copiously \cite{Amsler02}.

\section{Couplings to two photons}
The L3 and ALEPH collaborations at LEP have  reported new results in
$\gamma\gamma$ collisions around the $Z^0$ pole and at higher energies \cite{Acciarri,Barate}. The
$f_2(1270)$ is seen by ALEPH in the $\gamma\gamma\to\pi^+\pi^-$ mass spectrum, but
neither
$f_0(1500)$ nor
$f_0(1710)$ are observed and 95\% confidence level upper limits of 310 eV, respectively 550
eV are given for the formation and decay of these states into $\pi^+\pi^-$. According to
Crystal Barrel, 
$4\pi$ decays account for more than half of the decay rate of
$f_0(1500)$. The $\pi^+\pi^-$ decay branching
ratio was measured: a compilation
from early data \cite{RMP} gives 0.193 $\pm$ 0.050, in   agreement with a more 
recent 0.226 $\pm$ 0.079 from an extensive analysis of $\pbarp$ annihilation into 
$5\pi$ \cite{Abele01}. Using the latter and combining with ALEPH we therefore obtain for
the upper limit
\begin{equation}
\Gamma(f_0(1500) \to \gamma\gamma) \leq 1.4 \ {\rm keV}
\label{ulimit}
\end{equation}
at the 95 \% confidence level.

The L3 collaboration measuring $\gamma\gamma\to K_sK_s$ observes the $f_2(1270)$
and the $a_2(1320)$, the $f'_2(1525)$, and a dominantly spin 2 signal around 1750 MeV, which
is assigned to the $f_J(1710)$ (with, however, $J=2$). The measured partial width is
\begin{equation} 
\Gamma(f_2(1710) \to \gamma\gamma)\times \Gamma(f_2(1710) \to \KKbar)/\Gamma_{\rm tot} = 49
\pm 17 \ {\rm eV} \ .
\label{a2}
\end{equation}
However, roughly 25 \% of the signal is due to $J=0$. Assuming
this signal to come from the $f_0(1710)$ one obtains \cite{Braccini}
\begin{equation} 
\Gamma(f_0(1710) \to \gamma\gamma)\times \Gamma(f_2(1710) \to \KKbar)/\Gamma_{\rm tot} = 130
\pm 96 \ {\rm eV} \ .
\label{f0}
\end{equation}
A spin 0 signal at
1500 MeV is totally excluded by the L3 data, but the corresponding, presumably very small,
upper limit for $f_0(1500)$ is not given. We shall argue below that the spin 2 component could
instead be  due to the (isovector) radial excitation $a_2(1700)$ and that the spin 0 component
is consistent with a mainly $\ssbar$ $f_0(1710)$. 

\begin{figure}[htb]
\parbox{140mm}{\mbox{
\includegraphics[width=70mm]{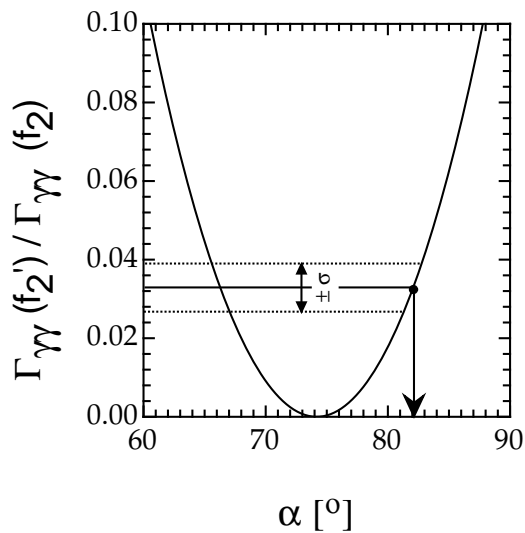}
}\centering}\hfill
\caption[]{Predicted ratio of two-photon widths for the $f_2(1270)$ and
$f_2'(1525)$ mesons as a function of $\alpha$. Note the twofold ambiguity.
The measured value of $\alpha$ consistent with the mass formulae is shown by the arrow.
\label{f2poverf2}}
\end{figure}

The two-photon width of  the isoscalar meson $f'$ (eqn. (\ref{psi})) with mass $m$ is
given within SU(3) by
\begin{equation}
\Gamma_{\gamma\gamma} = c\  (5 \cos\alpha - \sqrt{2} \sin\alpha)^2\  m^3 \ ,
\label{twogam}
\end{equation}
which vanishes for $\alpha$ = 74$^\circ$, and where $c$ is a nonet constant (for the isoscalar partner
$f$ in the nonet replace $\alpha$ by
$\alpha$ + 90$^\circ$). This formula is quite
reliable as it reproduces  the measured two-photon widths of the $2^{++}$ mesons $f_2(1270)$
and
$f_2'(1525)$ and those of the
$0^{-+}$ isoscalars  $\eta$ and $\eta'$ (the latter albeit with   10\% SU(3) violating
corrections). To gain confidence in the formula, let us examine the predicted ratio
$\Gamma_{\gamma\gamma}(f_2')/$ $\Gamma_{\gamma\gamma}(f_2)$ as a function of $\alpha$. This
is shown in fig.
\ref{f2poverf2} and compared to the measured ratio of 0.0328 $\pm$ 0.0062
\cite{Groom,Acciarri}. This ratio is extremely sensitive to the tensor mixing angle and one
obtains the accurate value
$\theta$ = (27.3 $\pm$ 0.8) $^\circ$, in excellent agreement with the mass formulae.

Also, we find  from eqn. (\ref{twogam}) and the two-photon partial width
$\Gamma_{\gamma\gamma}$ = 9  $c$  $m^3$ for the isovector meson $a$ the well-known ratios of
partial widths which read apart from
$m^3$ factors
\begin{equation}
\Gamma_{\gamma\gamma}(a) : \Gamma_{\gamma\gamma}(f) : \Gamma_{\gamma\gamma} (f') = 9 : 25 : 2
\ ,
\end{equation}
and which hold in the case of ideal mixing. Thus, as will become crucial in the following
discussion, a pure
$\ssbar$ isoscalar has a two-photon width an order of magnitude smaller than its $\nnbar$ partner.

The constant $c$
for the tensor nonet can be calculated from e.g. the two-photon width of
$f_2'(1525)$ \cite{Acciarri}: $c$ = 4.9 $\times 10^{-14}$ MeV$^{-2}$ which, in turn, leads to
a predicted two-photon partial width of 1.0 keV for the $a_2(1320)$, in excellent agreement with the
measured value \cite{Groom}.

Let us now deal with scalar mesons. In a non-relativistic calculation the two-photon width is
given by \cite{Tumanov}
\begin{equation}
\Gamma_{\gamma\gamma} (0^{++}) = k\left(\frac{m_0}{m_2}\right)^3\Gamma_{\gamma\gamma}
(2^{++}) \ ,
\label{zerototwo}
\end{equation}
with obvious notations. Here the factor $k=15/4$ stands for spin multiplicities. How
reliable is this relation? It can be checked with recent data on the charmonium
states
$\chi_{c2}$ and
$\chi_{c0}$: the ratio of $\chi_{c0}$ to $\chi_{c2}$ two-photon widths is measured to be 6.7
$\pm$ 2.1 while eqn. (\ref{zerototwo}) predicts the somewhat smaller value of 3.3.  
Relativistic calculations lead to a smaller value of scalar two-photon widths with $k \simeq 2$ \cite{Li}. Assuming
that
$f_0(1370)$ is the mainly $\nnbar$ scalar, one then predicts with formula (\ref{zerototwo}) a
two-photon width of 5.7 keV, in  good agreement with analyses of
$\gamma\gamma\to\pi\pi$ \cite{Morgan}. 

\begin{figure}[htb]
\parbox{140mm}{\mbox{
\includegraphics[width=100mm]{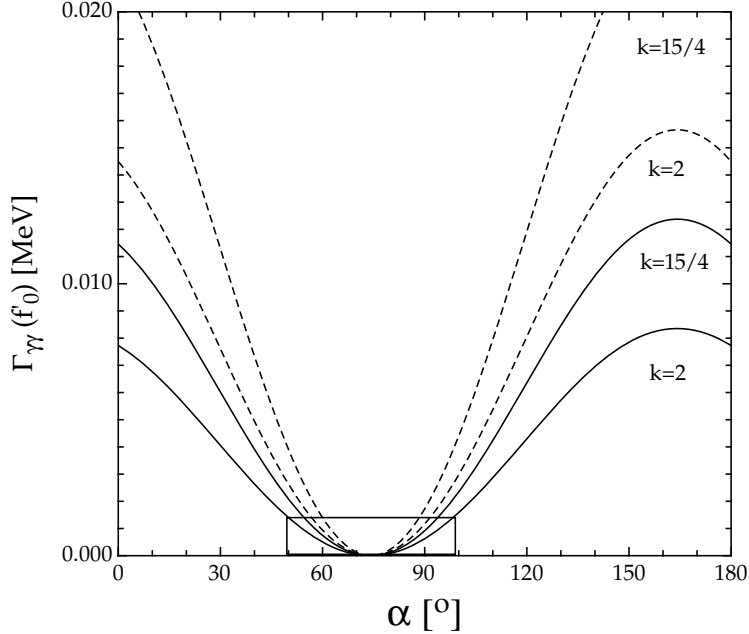}
}\centering}\hfill
\caption[]{Predicted two-photon width for the $f_0(1500)$ (full curves) and the $f_0(1710)$ (dashed
curves) assuming a $\qqbar$ structure. The box shows the 95\% confidence level upper limit from ALEPH
for the $f_0(1500)$.
\label{f0width}}
\end{figure} 

We shall therefore assume that for $k$ in the range  2
$\leq k \leq 15/4$ formula (\ref{zerototwo}) provides a reliable estimate of scalar two-photon widths.
Figure \ref{f0width} then shows the two-photon width expected for a $0^{++}$ isoscalar at 1500
MeV as a function of $\alpha$, calculated from eqns. (\ref{twogam}) and (\ref{zerototwo}),
together with the 95\% confidence level upper limit
(\ref{ulimit}) from ALEPH. Assuming a $\qqbar$ structure, the lack of $f_0(1500)$
production in
$\gamma\gamma$ implies  that this state must be mainly $\ssbar$ (50 $^\circ\leq\alpha\leq
100^\circ$). This is, however, in clear contradiction with the $\nnbar$ dominance discussed in
the previous section and therefore suggests that $f_0(1500)$ is of a different nature. 

Little
can be said about
$f_0(1710)$ from the ALEPH data, since its decay branching ratio into $\pi^+\pi^-$ is not
known. Assuming $\alpha$ $\sim$ 112 $^\circ$ from the previous section, we find
from fig.
\ref{f0width} and the ALEPH two-photon upper limit of 550 eV a $\pi^+\pi^-$ branching ratio of
at most 5 to 10 \%, hence small, as expected for a mainly $\ssbar$ state.

Let us now turn to the L3 data \cite{Acciarri} and to the signal observed around 1750 MeV.
Since decay branching ratios into $\KKbar$ are not known, only  qualitative statements can be
made. Assuming that the spin two component is due to the $a_2(1700)$ reported by L3 in
$\gamma\gamma\to 3\pi$ \cite{Acciarri97} and by Crystal Barrel in $\pbarp\to\pi^0\pi^0\eta$
\cite{Amsler02} one would obtain from the partial width (\ref{a2}) and with, say, a 5\% decay
branching ratio into $\KKbar$, a  two-photon width of $\simeq$ 1 keV. These numbers are
comparable to those for the ground state $a_2(1320)$. As far as the spin zero
signal is concerned, the upper limit  (\ref{f0}) can be accommodated by a nearly
pure $\ssbar$ with a large $\KKbar$ decay branching ratio (see fig. \ref{f0width}).

\section{Discussion and conclusions}
As we have argued, the measured $\pi\pi$, $\eta\eta$, $\eta\eta'$ and $\KKbar$ decay rates of
$f_0(1500)$ in $\pbarp$ annihilation and central production are consistent with those
expected for a
$\uubar+\ddbar$ state. On the other hand, $f_0(1500)$ is not produced in $\gamma\gamma$
collisions, with an upper limit at the level expected for a mainly $\ssbar$ state.
This contradiction suggests that $f_0(1500)$ is not a $\qqbar$ state. A four-quark state is unlikely since
its two-photon width is not significantly smaller than for a $\qqbar$ state. For example, a
partial width of $\simeq$ 0.6 keV is predicted in ref. \cite{Barnes} for the prominent candidate $a_0(980)$, while
formula (\ref{zerototwo}) gives  0.8 keV for a $\qqbar$ state (using 1 keV for the $a_2(1320)$ and $k=2$). 

The natural
explanation is that $f_0(1500)$ is a gluonium state, since the production of glue in
$\gamma\gamma$ collisions is suppressed. Its mass also lies in the correct range, lattice gauge calculations
predicting the ground state scalar glueball to be at 1611 $\pm$ 30 $\pm$ 160 MeV \cite{Michael}.

The $f_0(1710)$ appears to be dominantly $\ssbar$ from $\pi\pi$, $\eta\eta$ and $\KKbar$
data in central collisions. This is consistent with its absence in $\pbarp$
annihilation where scalars are copiously produced and no
mechanism is known preventing the production of  $\uubar+\ddbar$ scalars. We note that, surprisingly for an $\ssbar$
state, no signal for
$f_0(1710)$ was reported earlier in $K^-p\to K_SK_S\Lambda$  interactions \cite{Aston 88D}.
However, the assumption in the analysis was that its spin was two. The
$\ssbar$ dominance is consistent with (but not required by) $\gamma\gamma$ data, since only
few decay branching ratios have  been measured for this state. In particular, the absolute
two-body decay rates are not known and are urgently needed from future experiments. 

If one now assumes 
$a_0(1450)$, $f_0(1370)$, $f_0(1710)$ and $K_0^*(1430)$, one obtains from the linear
mass formula
\begin{equation}
\tan^2\theta = \frac{4m(K_0^*)-m(a_0)-3m[f_0(1710)]}{3m[f_0(1370)]+m(a_0)-4m(K_0^*)} 
\end{equation}
the $0^{++}$ mixing angle $\alpha = \theta$ + 54.7$^\circ  \simeq 114^\circ$ for a low mass (1300 MeV) $a_0$ or
$\simeq 127^\circ$ for a high mass  (1470 MeV) $a_0$ \cite{Groom}. We have assumed for $f_0(1370)$ a mass of 1360
MeV \cite{RMP}. The mixing angle is also sensitive to the mass
of this state which is not well known
\cite{Groom}. It is interesting to note that these values of $\alpha$ are in the
correct range, see the data on the $f_0(1710)$ in fig. \ref{etakkpipi} above.  

The glueball is the $f_0(1500)$ in our model. This may be too simple minded, since the two
$\qqbar$ isoscalars are likely to mix with the nearby glueball. Nonetheless these results
strongly favour the scenario where
$f_0(1500)$ is largely glue and
$f_0(1710)$ dominantly
$\ssbar$
\cite{FEC,Kirk}, and disfavour $f_0(1710)$ being the glueball \cite{Weingarten}. 
More quantitative statements will have to await more accurate $\gamma\gamma$ data and
theoretical guidance on the production rate of glueballs in electromagnetic processes.

\end{document}